\documentstyle[amsfonts,aps,prl,twocolumn,epsf]{revtex}

\newcommand{\beq}{\begin{equation}}
\newcommand{\eeq}{\end{equation}}
\newcommand{\beqa}{\begin{eqnarray}}
\newcommand{\eeqa}{\end{eqnarray}}

\begin{document}

\title{A thermal model for adaptive competition in a market}

\author{
Andrea Cavagna\thanks{E-mail: a.cavagna1@physics.ox.ac.uk},
Juan P. Garrahan\thanks{E-mail: j.garrahan1@physics.ox.ac.uk},
Irene Giardina\thanks{E-mail: i.giardina1@physics.ox.ac.uk}
and 
David Sherrington\thanks{E-mail: d.sherrington1@physics.ox.ac.uk}
}

\address{
Theoretical Physics, University of Oxford,
1 Keble Road, Oxford OX1 3NP, United Kingdom
}

\date{July 13, 1999}
\maketitle

\begin{abstract}
{{\bf Abstract:} 
\sl
New continuous and stochastic extensions of the minority game, 
devised as a fundamental model for a market of competitive agents, 
are introduced and studied in the context of statistical physics. 
The new formulation reproduces the 
key features of the original model, without the need
for some of its special assumptions and, most importantly, it demonstrates 
the crucial role of stochastic decision-making. 
Furthermore, 
this formulation provides the exact but novel non-linear equations for 
the dynamics of the system.
}
\end{abstract}

There is currently much interest in the statistical physics of
non-equilibrium frustrated and disordered many-body 
systems \cite{noneq}. 
Even relatively simple microscopic dynamical equations have been shown to
lead to complex co-operative behaviour. Although several of the
interesting examples are in areas traditionally viewed as physics, 
it is growingly apparent that many further challenges for statistical 
physics have their origins in other fields like biology \cite{bio} 
and economics \cite{pwa}. 
In this letter we discuss a simple model whose origin lies in a 
market scenario and show that not only does it exhibit interesting 
behaviour in its own right but also it yields an intriguingly
unusual type of stochastic micro-dynamics of potentially more general
interest.

The model we will introduce is based on the 
the minority game (MG) \cite{challet1},
which is a simple
and in\-tui\-tive model for the behaviour of a 
group of agents subject to the economic law of supply and demand,
which ensures that in a market the profitable group of buyers or
sellers
of a commodity is the minority one \cite{arthur}.
From the perspective of statistical physics, these problems 
are novel examples of frustrated and disordered many-body
systems. 
Agents do not interact directly but with their collective action
determine a `price' which in turn affects their future
behaviour, so that minority reward implies frustration.
Quenched disorder enters in
that different agents respond in different ways to the
same stimuli.
There are effective random interactions between 
agents via the common stimuli
and the cooperative behaviour 
is reminiscent of that of spin-glasses \cite{mpv},
but there are important conceptual and technical differences 
compared 
with the problems of conventional statistical physics. 

The setup of the MG in the original
formulation of \cite{challet1} is the following: 
$N$ agents choose at each time step whether 
to `buy' ($0$) or `sell' ($1$). 
Those agents who have made the minority choice win, 
the others lose. 
In order to decide what to do agents use 
strategies, which prescribe an action given the set 
of winning outcomes in the last $m$ time steps. 
At the beginning of the game each agent draws $s$ strategies 
randomly and keep them forever.
As they play, the agents give points 
to all their strategies according to their potential success in 
the past, and at 
each time step they employ their currently most successful one 
(i.e. the one with the highest number of points). 

The most interesting macroscopic observable in the MG is the
fluctuation $\sigma$ of the excess of buyers to sellers.
This quantity is equivalent to the price volatility in 
a financial context and it is a measure of the global waste 
of resources by the community of the agents. We therefore 
want $\sigma$ to be as low as possible.
An important feature of the MG, observed in simulations \cite{savit1}, 
is that there is a regime of the parameters where $\sigma$ is 
{\it smaller} than 
the value $\sigma_r$ which corresponds to the 
case where each agent is buying or selling randomly. 
Previous studies have considered this feature from a geometrical and
phenomenological point of view
\cite{challet2}.
Our aim, however, is to enable a full analytic solution.

One of the major obstacles to an analytic study of the MG in 
its original formulation is the presence of an explicit time
feedback via the memory $m$. Indeed, when 
the information processed at each time step by the agents is 
the {\it true} history, 
that is the result of the choices of the agents in the $m$ previous
steps, the dynamical evolution of the system is non-Markovian and 
an analytic approach to the problem is very difficult.

A step forward in the simplification of the
model has been made in \cite{mino}, where it has been 
shown that the explicit memory of the agents is actually irrelevant
for the global behaviour of the system: when
the information processed by the agents at each time step is 
just {\it invented} randomly, having nothing to do with the true time series, 
the relevant macroscopic observables do not change. 
The significance of this result is the following: the crucial
ingredient for the volatility
to be reduced below the random value appears to be that the agents 
must all react to the {\it same} piece of information, 
irrespective of whether this information is true or false
\cite{sunspot}.
This result has an important technical 
consequence, since the explicit time feedback introduced by the 
memory disappears: the agents respond now to an instantaneous
random piece of information, 
i.e. a noise, so that the process has become stochastic and Markovian. 

The model can be usefully simplified even further and at the same
time generalized and made more realistic. Let us first consider  the
binary nature of the original MG. It is clear that from a simulational
point of view a binary setup offers advantages of computational efficiency,
but unfortunately it is less ideally suited for an analytic 
approach \cite{cm}. 
More specifically, if we are interested in the analysis of 
time evolution, integer variables are usually harder to handle. 
Moreover, the geometrical considerations that have been made
on a hypercube of strategies of dimension $2^m$ for the binary setup
\cite{challet2}, become more natural and 
general if the strategy space is continuous.
Finally, in the original binary formulation of the MG there is 
no possibility for the agents to fine tune their bids: each
agent can choose to buy or sell, 
but they cannot choose by {\it how much}.
As a consequence, also the win or loss of the agents is not
related to the consistency of their bids. This is another 
unrealistic 
feature of the model, which can be improved. For all these reasons, 
we shall now introduce a continuous formulation of the MG.

Let us define a strategy $\vec R$ as a vector in the real space
${\mathbb R}^D$, subject to the constraint, $\| \vec R \| = \sqrt{D}.$
In this way the space of strategies $\Gamma$ is just a 
sphere and strategies can be thought as points on it.
The next ingredient we need is the information processed by
strategies. To this aim we introduce a random noise $\vec \eta(t)$, 
defined as a unit-length vector in ${\mathbb R}^D$,
which is $\delta$-correlated in time
and uniformly distributed on the unit sphere. 
Finally, we define the response $b(\vec R)$ of a strategy 
$\vec R$ to the information $\vec \eta(t)$, as the projection of 
the strategy on the information itself,
\beq
b(\vec R) \equiv \vec R\cdot \vec \eta(t) \ .
	\label{bid}
\eeq
This response is nothing else than the {\it bid} prescribed
by the particular strategy $\vec R$. The bid is now a continuous 
quantity, which can be positive (buy) or negative (sell). 

At the beginning of the game each agent draws $s$ strategies randomly 
from $\Gamma$, with a flat distribution. All the strategies initially
have zero points and in operation the points are updated in a manner 
discussed below.
At each time step the agent uses his/her strategy with the highest number
of points. The {\it total bid} is:
\beq
A(t) \equiv \sum_{i=1}^N b_i(t) 
= \sum_{i=1}^N \vec R^\star_i(t)\cdot 
\vec\eta(t) \ ,
	\label{tot}
\eeq
where $\vec R^\star_i(t)$ is the best strategy (that with the 
highest number of points) of agent $i$ at time
$t$. 

We have now to update the points. This is particularly simple in
the present continuous formulation. Let us introduce a time dependent 
function $P(\vec R,t)$ defined on
$\Gamma$, which represents the points $P$ of strategy $\vec R$
at time $t$.
We can write a very simple and intuitive time evolution equation for 
$P$,
\beq
P(\vec R,t+1) = P(\vec R,t) - A(t) \, b(\vec R) / N \ ,
	\label{evol}
\eeq
where $A(t)$ is given by eq.(\ref{tot}). 
A strategy $\vec R$ is thus rewarded (penalized) if its bid has an opposite 
(equal) sign to the total bid $A(t)$, 
as the supply-demand dynamics requires. 
Now the win or the loss is proportional to the bid. 
\begin{figure}
\begin{center}
\leavevmode
\epsfxsize=3in
\epsffile{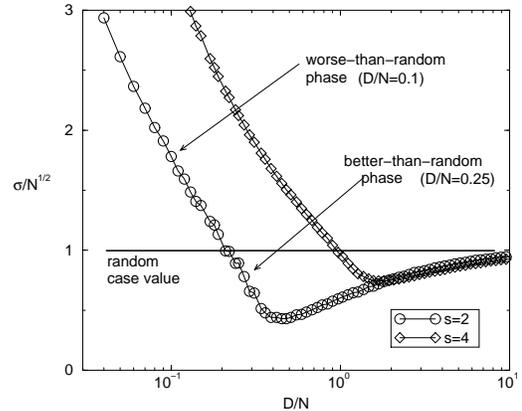}
\caption{Continuous formulation: The scaled variance $\sigma/\sqrt{N}$
as a function of the reduced dimension $D/N$, 
at $s=2$ and $s=4$. The horizontal line 
is the variance in the random case. The total time $t$ 
and the initial time $t_0$ are $10000$ steps. Average over $100$
samples, $N=100$.
}
\label{fig1}
\end{center}
\end{figure}
It is important to check whether the results obtained with this 
continuous formulation of the MG are the same as in the original binary
model. The main observable of interest is the 
variance (or volatility) $\sigma$ in the 
fluctuation of $A$, $\sigma^2 = \lim_{t\to\infty} \frac{1}{t} 
\int_{t_0}^{t_0+t} dt' \; A(t')^2$. Indeed, we shall not
consider any quantity related to individual agents.
We prefer to concentrate on the global behaviour of the system, 
taking more the role of the market regulator than that of a
trading agent.
The main features of the MG are reproduced: first, we have checked 
that the relevant scaling parameter is the reduced dimension
of the strategy space $d=D/N$; 
second, there is a regime of $d$
where the variance $\sigma$ is smaller than the random value 
$\sigma_r$, showing a minimum at $d=d_c(s)$,  
and, moreover, the minimum of $\sigma(d)$ is shallower the higher
is $s$ \cite{challet2}; see Fig.1.
It can be shown that {\it all} the other 
standard features of the binary model are reproduced in the 
continuous formulation.

An interesting observation is that 
there is no need 
for $\vec\eta(t)$ to be random at all. Indeed, the only requirement is that 
it must be {\it ergodic}, spanning the whole space $\Gamma$,
even in a deterministic 
way. Moreover, if $\vec\eta(t)$ visits just a sub-space of 
$\Gamma$ of dimension
$D'<D$ everything in the system proceeds as if 
the actual dimension was $D'$: the {\it effective} 
dimension of the strategy space is fixed by the dimension of the 
space spanned by the information.

Relations (\ref{tot}) and (\ref{evol}) constitute  
a closed set of equations for the dynamical evolution of 
$P(\vec R,t)$, whose solution, once averaged over 
$\vec \eta$ and over the initial distribution of the strategies, 
gives in principle an exact determination of the behaviour of the 
system. In practice, the presence of the `best-strategy' rule, i.e.
the fact that each agent uses the strategy with the highest points,
makes the handling of these equations still difficult.
From the perspective of statistical physics it is natural 
to modify the deterministic
nature of the above procedure by introducing a thermal description
which progressively allows stochastic deviations from the `best-strategy'
rule, as a temperature is raised. 
We shall see that this generalization is also advantageous,
both for the performance of the system in certain regimes and for the
development of convenient analytical equations for the dynamics. 
In this context the 
`best-strategy' original formulation of the MG can be viewed as a 
zero temperature limit of a more general model.

Hence we introduce the Thermal Minority Game (TMG), defined 
in the following way. 
We allow each agent a certain degree of stochasticity in the
choice of the strategy to use at any time step.
For each agent $i$ the probabilities of employing his/her strategy 
$a=1,\dots,s$ is given by,
\beq
\pi^a_i(t)\equiv \frac{e^{\beta P(\vec R_i^a,t)}}{Z_i} \  , \ \ \ \ 
Z_i\equiv \sum_{b=1}^s e^{\beta P(\vec R_i^b,t)} \ ,
	\label{proba}
\eeq
where $P$ are the points, evolving with eq.(\ref{evol}).
The inverse temperature $\beta=1/T$ 
is a measure of the power of {\it resolution} of the agents: 
when $\beta\to\infty$ they
are perfectly able to distinguish which is their best strategy, while
for decreasing $\beta$ they are more and more confused, 
until for $\beta=0$ they choose their strategy completely at random. 
What we have defined is therefore a model which interpolates 
between the original `best-strategy' MG ($T=0$, $\beta=\infty$) and 
the random case ($T=\infty$, $\beta=0$). 
In the language of Game Theory, when $T=0$ agents play `pure'
strategies, while at $T>0$ they play `mixed' ones \cite{games}.

We now consider the consequences of having introduced the 
temperature. 
Let us fix a value of $d$ belonging to the worse-than-random phase of 
the MG (see Fig.1) and see what happens to 
the variance $\sigma$ when we switch on the temperature. 
We do know that for $T=0$ we must recover the same value as 
in the ordinary MG, while for $T\to\infty$ we must obtain the
value $\sigma_r$ of the random case. But in the middle 
a very interesting thing occurs:
$\sigma(T)$ is not a monotonically decreasing function of $T$, but there 
is a large intermediate temperature regime where $\sigma$ is {\it smaller} 
than the random value $\sigma_r$. This behaviour is shown in Fig.2.

The meaning of this result is the following:
even if the system is in a MG phase which is worse than random, 
there is a way to significantly decrease the volatility $\sigma$ below
the random value $\sigma_r$ by {\it not} always 
using the best strategy, but rather allowing a certain degree of 
individual error.
\begin{figure}
\begin{center}
\leavevmode
\epsfxsize=3in
\epsffile{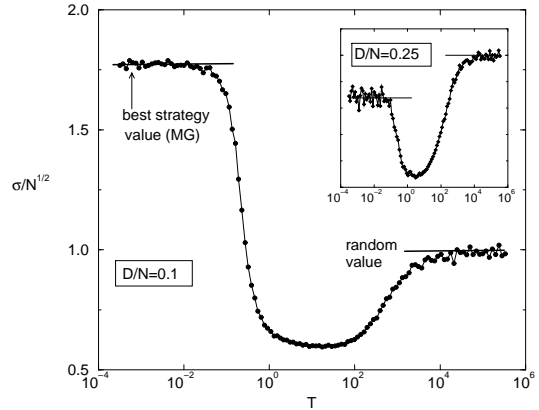}
\caption{TMG: The scaled variance $\sigma/\sqrt{N}$ as a
function of the temperature $T$, at $D/N=0.1$, for
$s=2$. In the inset we show $\sigma(T)/\sqrt{N}$
for $D/N = 0.25$.}
\label{fig2}
\end{center}
\end{figure}
Note from Fig.2 that the temperature range where the variance is smaller
than the random one is more than two orders of magnitude large, 
meaning that almost every kind of 
individual stochasticity of the agents improves the global behaviour 
of the system.
Furthermore, as we show in the inset of Fig.2, 
if we fix $d$ at a value belonging
to the better-than-random phase, but with $d<d_c$, a similar range
of temperature
still improves the behaviour of the system, decreasing the volatility 
even below the MG value.
\begin{figure}
\begin{center}
\leavevmode
\epsfxsize=3in
\epsffile{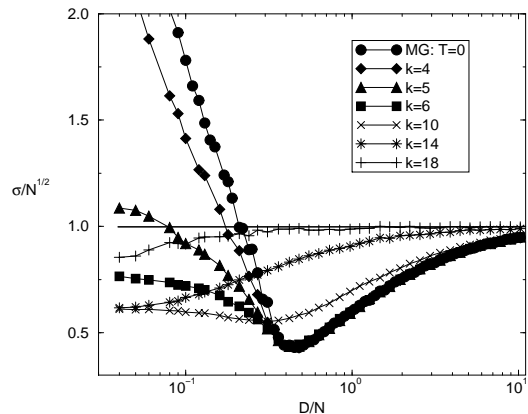}
\caption{TMG: The scaled variance $\sigma/\sqrt{N}$ as a function 
of the reduced dimension $D/N$, 
at different values of the temperature $T=2^k \times 10^{-2}$, for
$s=2$.}
\label{fig3}
\end{center}
\end{figure}
These features 
can be seen also in Fig.3, where we plot $\sigma$ as 
a function of $d$ at various values of the temperature. 
In addition this figure shows further effects:
(i) the improvement due to thermal noise occurs only for $d<d_c$;
(ii) there is a cross-over temperature $T_1 \sim 1$, 
below which temperature has
very little effect for $d>d_c$; (iii) above $T_1$
the optimal $d_c(T)$ moves continuously towards zero
and $\sigma(d_c)$ increases; (iv) there is a
higher critical temperature
$T_2 \sim 10^2$ at which $d_c$ vanishes, and for
$T>T_2$ the volatility
becomes monotonically increasing with $d$.

We turn now to a more formal description of the TMG.
Once we have introduced the probabilities $\pi_i^a$ in eq.(\ref{proba}) 
we can write a dynamical equation for them. Indeed, from
eq.(\ref{evol}), after taking the continuous-time limit, we have,
\beq
\dot\pi_i^a(t)= 
-\beta \; \pi_i^a(t) \; a(t)
\left( \vec R_i^a - 
	\sum_{b=1}^s \pi_i^b(t) \; \vec R_i^b \right)\cdot
 	\vec\eta(t) \ ,
	\label{snarpo}
\eeq
where the normalized total bid $a(t)$ is given by,
\beq
a(t)= N^{-1} \, \sum_{i=1}^N \vec r_i(t) \cdot \vec \eta(t) \ .
	\label{tre}
\eeq
Now $\vec r_i(t)$ is a stochastic variable, drawn at each time $t$
with the time dependent probabilities set $[\pi_i^1,\dots,\pi_i^s]$.
Note the different notation: $\vec R_i^a$ are the {\it quenched}
strategies, while $\vec r_i(t)$ is the particular strategy 
drawn at time $t$ from the set $[\vec R^1_i,\dots,\vec R^s_i]$
by agent $i$ with instantaneous probabilities 
$[\pi^1_i(t),\dots,\pi^s_i(t)]$. 
In order to better understand equation (\ref{snarpo}), we recall
that $b_i^a(t)=\vec R_i^a\cdot\vec\eta(t)$ is the bid of 
strategy $\vec R_i^a$ at time $t$ (eq.(\ref{bid})) and therefore the quantity 
$w_i^a(t)\equiv - B(t) b_i^a(t)$ can be considered as the {\it win} of 
this strategy (cf. eq.(\ref{evol})). 
Hence, we can rewrite eq. (\ref{snarpo}) in the following more 
intuitive form,
\beq
\dot\pi_i^a(t)= \beta \; \pi_i^a(t) \;
[\; w^a_i(t)- \langle w\rangle_i \; ] \ ,
	\label{fine}
\eeq
where $\langle w\rangle_i \equiv \sum_{b=1}^s \pi_i^b(t) \; w^b_i(t)$. 
The meaning of equation (\ref{fine}) is clear: 
the probability $\pi^a_i$ of a particular stra\-te\-gy $\vec R^a_i$
increases only if the performance of that strategy is better  
than the instantaneous {\it average} performance of all the strategies 
belonging to the same agent $i$ with the same actual
total bid \cite{re}.

Relations (\ref{snarpo}) and (\ref{tre}) are the exact dynamical 
equations for the TMG. 
They do not involve points nor memory, 
but just stochastic noise and quenched 
disorder, and they are local in time. 
From the perspective of statistical mechanics, 
this is satisfying and encouraging. 
However, these equations differ
fundamentally from conventional replicator and Langevin
dynamics. First, the Markov-propagating variables are themselves 
probabilities. Second, there are two sorts of stochastic noises, 
as well as quenched randomness. Third, and more importantly, 
the stochastic noises enter non-linearly, one independently
for each agent
via probabilistic dependence on the $\pi$
themselves, 
the other globally and quadratically.
They thus provide interesting challenges for
fundamental transfer from microscopic to macroscopic dynamics, 
including an identification 
of the complete set of necessary order parameters \cite{cls}. 
We shall address 
the problem of finding a solution of the TMG equations in a future work.

Finally, let us note that the TMG (as well as the MG) is not only 
suitable for the description of market dynamics.  Indeed, any natural 
system where a population of individuals must organize itself in order to
optimize the utilization of some limited resources, is qualitatively 
well described by such a model. 
We hope that the thermal model we have introduced in this Letter 
will give more insight into this kind of natural phenomena.

We thank L. Billi, P. Gillin and P. Love for many suggestions, and  
N.F. Johnson and I. Kogan for useful discussions. This work was
supported by EPSRC Grant GR/M04426 and EC Grant ARG/B7-3011/94/27.

\end{document}